\documentclass[3p]{elsarticle}
\usepackage{amsmath}
\usepackage{hyperref}

\journal{Physica A}

\begin{document}

\begin{frontmatter}

\title{Prisoner's dilemma game on complex networks with a death process:
Effects of minimum requirements and immigration}

\author[mymainaddress]{Wonhee Jeong}
\author[mymainaddress]{Unjong Yu\corref{mycorrespondingauthor}}
\cortext[mycorrespondingauthor]{Corresponding author}
\ead{uyu@gist.ac.kr}
\address[mymainaddress]{Department of Physics and Photon Science, Gwangju Institute of Science and Technology, Gwangju 61005, South Korea}

\begin{abstract}
We present results of the prisoner's dilemma game on complex networks that have population change. We introduce a death process with minimum requirements and show that it induces a highly cooperative society. We also study the effects of immigration on the society. We show that the acceptable number of immigrants of the society is determined by the population of the society, the ratio of cooperator among immigrants, and the immigration interval. In addition, if immigrants have a preferential attachment link, the acceptable number of immigrants increases.
\end{abstract}


\end{frontmatter}


\section{Introduction}

A human cannot live alone without the help of others and has to make a relationship with people based on mutual cooperation. This network of cooperation is a foundation of the society. However, ironically, the human is tempted to act selfishly in the society \cite{Hardin1243,RAND2013413}. When others act cooperatively, a selfish human may gain more benefits than a cooperative human because the selfish human gains cooperation without investments. In reality, the selfish behavior can be observed often in the society. A human has contradictory aspects: cooperation and selfishness. It is a challenge to explain this contradiction of human cooperation, and many pioneers found some conditions that cooperation occurs \cite{rockenbach2006efficient,milinski2002reputation,rand2011dynamic}. However, the answer is not clear yet.

Recently, the evolutionary game theory has been used to explain how cooperation occurs \cite{nowak2006evolutionary,PERC20171}. The evolutionary game theory has two major concepts. The first one is a game. Among many kinds of games, prisoner's dilemma (PD) game well expresses human selfishness and has been studied a lot \cite{rand2011dynamic,wang2017onymity,gallo2015effects}. In this paper, we also use the PD game. The other major concept is the imitation process. In the evolutionary game theory, each node imitates neighbor's strategy following a given rule \cite{nowak1992evolutionary,PhysRevLett.98.108103,santos2006evolutionary}. However, the studies in this field have focused largely on the strategy spreading without population change. In this paper, we study the effect of population change by including a death process and immigration. The population may decrease with the death process based on the minimum requirements rule.

Humans have minimum requirements to survive. If a human does not satisfy minimum requirements, the human is culled (dies out) \cite{gleick1996basic,leslie1984caloric}. For this reason, we set the minimum requirements as the rule of the death process. In our simulation, if an individual does not satisfy minimum requirements in the previous game, the individual dies out with a certain probability; it means that the individual cannot adapt in the society. As a result of the death process with minimum requirements, a highly cooperative society is induced. This is one possible answer to why humans act cooperatively against others.

A highly cooperative society looks stable and good for meeting minimum requirements. Thus, outsiders would want to live in the society, and they may immigrate into the society \cite{massey1990social}. Although this is happening in the real world, immigration occurs restrictively to avoid the disorder of the society \cite{goldin1994political}. To study why restrictions are needed, we immigrate nodes in the highly cooperative society and observe the population change of the society. We show that the maximum number of immigrants the society can accept is influenced by the conditions of the society and the way of immigration of immigrants.

\section{Model and methods}

We model the society using the concept of the network.
Node and edge of the network represent individual and relation between the two individuals, respectively.
Individuals in a network interact only with individuals connected by edges.
We begin the simulation from the Erd\H{o}s-R\'{e}nyi (ER) random network \cite{erdos1959random}. ER random network is simple and has been used in many evolutionary dynamics studies \cite{PhysRevLett.98.108103,PhysRevE.79.016107,PhysRevE.80.026105}. Initially, each node selects its strategy randomly. After that, at each time step, we implement the PD game, imitation process, and death process sequentially.

In the PD game, each member can choose one strategy, cooperation ($C$) or defection ($D$). Thus, there are four possible cases: $CC$, $CD$, $DC$, and $DD$. In each case, payoffs that the individual gains are expressed in the payoff matrix \cite{nowak1992evolutionary,PhysRevLett.98.108103},
\begin{center}
\begin{tabular}{c|c c}
 & $C$ & $D$ \\
\hline
$C$ & $1$ & $0$ \\
$D$ & $b$ & $0$ \\
\end{tabular}
\end{center}
Here, the temptation of selfish behavior $b$ satisfies $1<b<2$. This is a weak version of prisoner's dilemma game and is used broadly to analyze strategy transitions according to temptation \cite{nowak1992evolutionary,PhysRevLett.98.108103}.

In the imitation process, imitation probability is given by \cite{PhysRevLett.98.108103,santos2006evolutionary},
\begin{equation}
 P_{i \rightarrow j} = \frac{\pi_j-\pi_i}{b \times \mathrm{max}\{k_i,k_j\}} ,\, \mathrm{when}\; \pi_j > \pi_i .\label{E1}
\end{equation}
$P_{i \rightarrow j}$ is the probability that node $i$ imitates the strategy of node $j$. Node $j$ is selected randomly among neighbors of node $i$. $\pi_i$ is $i$'s payoff in the previous round, and $k_i$ means the degree of node $i$. If $\pi_j$ is less than $\pi_i$, node $i$ keeps its strategy. 
Every node has a chance to change its strategy at each time step; after all the node have decided their next strategies, they update their strategies at the same time synchronously.

In the death process with minimum requirements $\pi_r$, if $\pi_i < \pi_r$, then node $i$ dies out with the death probability $P_d$. This process may reduce the population, which is the number of nodes in the network.

We also studied effects of immigration (see \cite{Tarik}, and references therein). A group of nodes are added in the network when the network reaches an equilibrium state. Here, we call the action of adding nodes ``immigration'' and an added node is called ``immigrant'', and a node that lived in the society before immigration is called ``native''. 
The immigration in ordinary condition is usually on a small scale and very selective, and gives only minor influence on the society. To the contrary, the immigration initiated by special events such as economic crisis and war causes many immigrants. Since this kind of the immigration is abrupt, it causes more serious and critical change to the society \cite{jofre2016immigration,kirisci_1995}. Therefore, we focus on this massive immigration in this paper. We fix the number of immigrants at one time. For example, if $200$ immigrants enter the society and the number of immigrants at one time is $50$, then the total number of immigration is $4$. The time interval between immigration is the immigration interval. At the end of all immigration, we simulate $20000$ time steps more for adaptation of immigrants. The rule of minimum requirements applies equally to immigrants. When the immigrants enter the society, they make links with some members of the society. In our simulation, the number of links of an immigrant is determined within the range of $\pm 1$ from $\langle k \rangle$ in the society. (However, if the number of links of immigrants is determined by the degree distribution of the society, the properties that we will show do not change.) Immigrants connect randomly to members of the society.

When immigrants enter the society, the population of the society changes. To quantify population change posed by the immigration, we define the change of population of the society by a unit immigrant $\chi$ as
\begin{equation} \notag
\chi \equiv \frac{\mathrm{Population\; change\; of\; the\; society}}{\mathrm{Total\; number\; of\; immigrants}}.
\end{equation}
If $\chi= 1$, then all immigrants adapt to the society without the death of natives. On the contrary, if $\chi = -1$, then the population of the society is reduced by the number of immigrants in spite of the immigration. If $\chi=0$, the population of the society does not change. In this paper, we say immigration is successful when $\chi>0$. Also, we control three variables to observe the properties of the massive immigration. The first one is $R_i$, which is the total number of immigrants with respect to the number of natives before immigration occurs ($N$). The second one is the ratio of cooperator among immigrants at each time ($R_c$). The last one is the immigration interval.

\section{Results and discussion}

\begin{figure}
\centering
\includegraphics[angle=270,width=1\columnwidth]{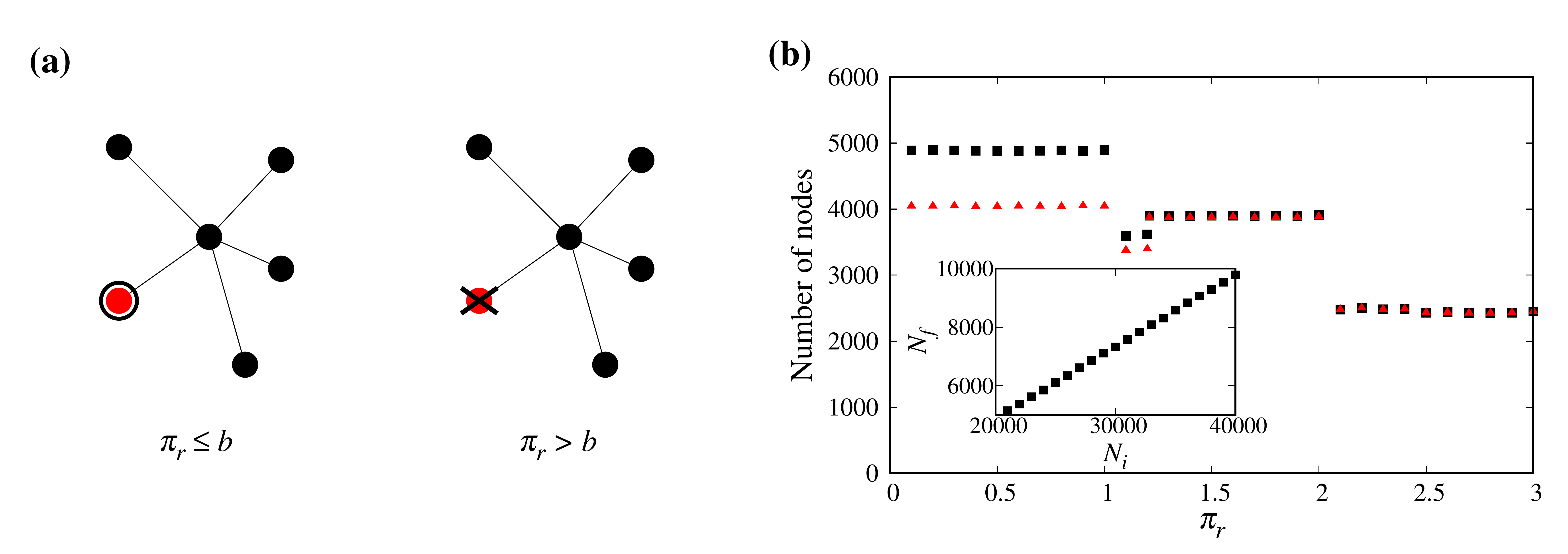}
\caption{(a)~Black and red circles represent cooperators and defectors, respectively. If $\pi_r \leq b$ (left), then the defector satisfies minimum requirements. However, if $\pi_r > b$ (right), then the defector dies out with probability $P_d$. (b)~The number of total live nodes (square) and the number of live nodes with cooperative strategy (triangle) as a function of minimum requirements $\pi_r$. In this simulation, we set $b=1.2$ and the number of nodes of the initial network $N_i$ is $10000$. At each simulation, we simulated until the system was in equilibrium. (At least $20000$ time steps were simulated.) $1000$ simulations were averaged in each case. Introduction of minimum requirements induces highly cooperative society through the death process.
(inset)~The number of live nodes ($N_f$) as a function of the number of initial nodes ($N_i$). In this simulation, we set $b=1.2$ and $\pi_r = 2.5$. $500$ simulations were averaged in each case.}
\label{fig1}
\end{figure}

In order to verify the effect of minimum requirements, we made ER random network of $\langle k \rangle = 12$ and performed the simulation with $P_d = 0.1$ and $b = 1.2$. Figure~\ref{fig1}(b) shows the number of live nodes and the number of live nodes with cooperative strategy as a function of minimum requirements. Since a death process reduces the number of nodes, the number of nodes in the final network is less than the initial network. The population of the final network has a linear dependence on that of the initial network as shown in the inset of Fig.~\ref{fig1}(b). Thus, if we start with more initial nodes, we get a larger number of live nodes in the final network. Note that the cooperation level of the final network is very high even with very small $\pi_r$. It is because defector core (defector surrounded by defectors) does not gain any payoff and so its payoff is always less than $\pi_r$. Thus, defector cores die out with $P_d$ by the death process.

The results of this simulation have an untrivial point, $\pi_r = b$. When $\pi_r \leq b$, if a defector has at least one cooperative neighbor, then the defector has no probability to die out. Thus, in this condition, the final network is not purely cooperative. However, when $\pi_r > b$, defectors who have only one cooperative neighbor die out with $P_d$; defectors need at least two cooperative neighbors. (See Fig.~\ref{fig1}(a).) In this condition, the final network is mostly purely cooperative, though it is not always purely cooperative. When the final network is purely cooperative, all nodes are cooperator cores (cooperators surrounded by cooperators).

\begin{figure}
\centering
\includegraphics[angle=270,width=1\columnwidth]{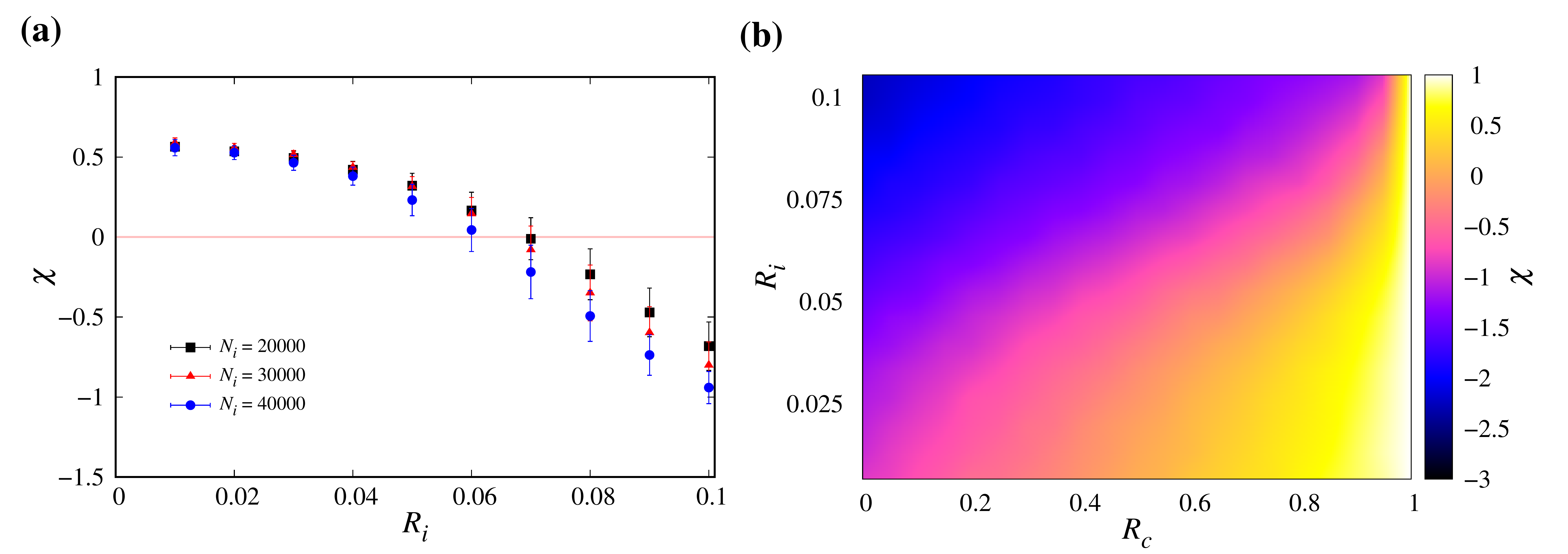}
\caption{We set $b=1.2$ and $\pi_r = 2.5$, and the immigration interval is $1000$. (a)~$\chi$ as a function of $R_i$ for each $N_i$. In this simulation, $R_c$ is fixed by $0.8$. $1000$ simulations were averaged in each society and $10$ simulations were averaged among the same $N_i$ in each point. (b)~We make highly cooperative society that the population is $N=7582$ and observe $\chi$ with varying $R_c$ and $R_i$. $1000$ simulations were averaged in each case. When the number of immigrants is small and the ratio of cooperator of immigrants is high, $\chi$ is larger than zero. That is, the population of the society becomes larger than before.}\label{fig2}
\end{figure}

\begin{figure}
\centering
\includegraphics[angle=270,width=0.46\columnwidth]{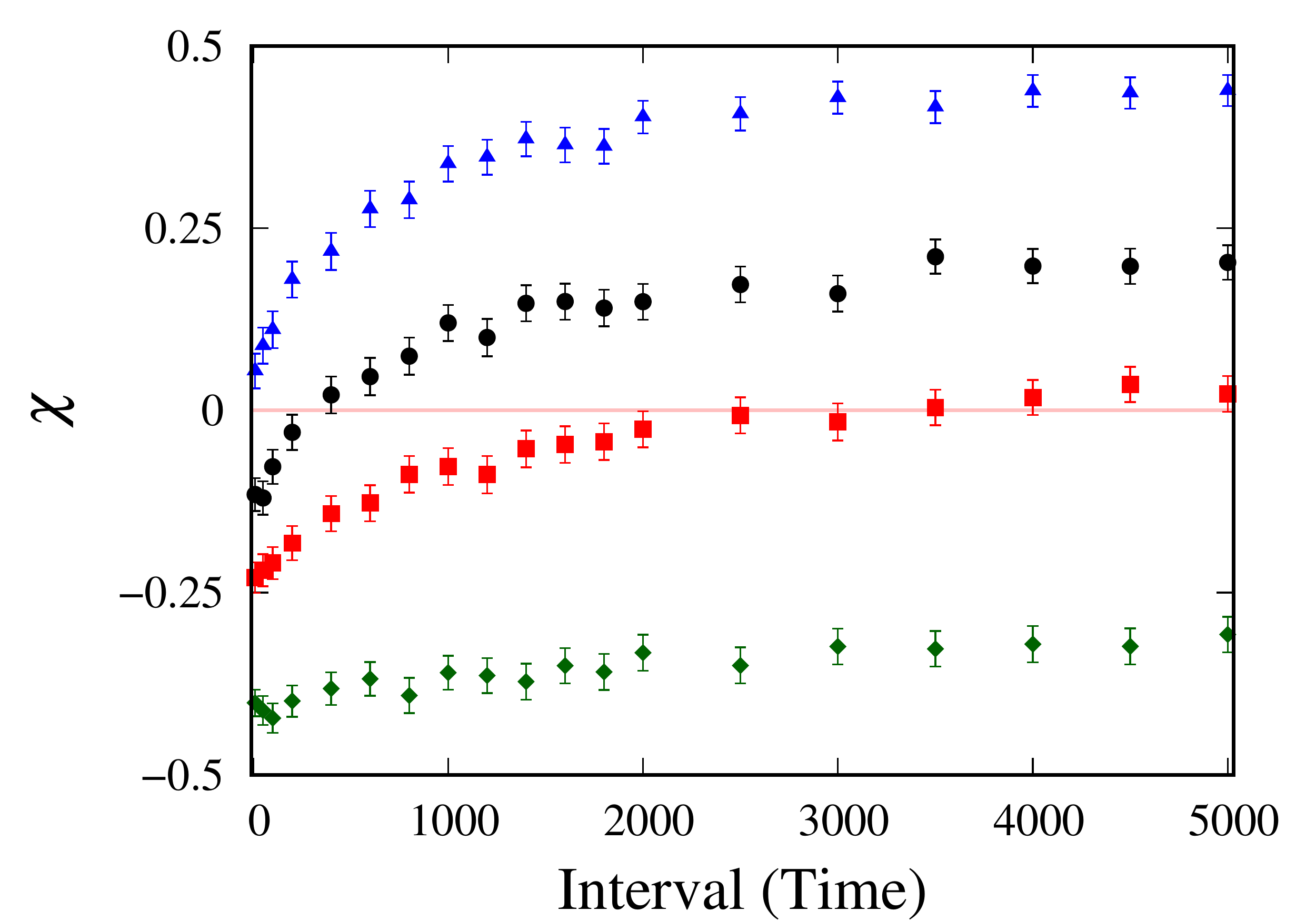}
\caption{In the same society as in Fig.~2(b), we fix the number of total immigrants to be $450$ ($R_i \approx 0.06$) and observe $\chi$ as a function of the immigration interval for each $R_c$. Values of $R_c$ are $0.9$, $0.85$, $0.8$, and $0.7$, from top to bottom. $3000$ simulations were averaged in each case.}\label{fig3}
\end{figure}

To identifying the properties of the immigration, we firstly set $b=1.2$, $\pi_r = 2.5$, and the immigration interval to be $1000$ in the highly cooperative society. From now on, these values do not change except in Fig.~\ref{fig3}, where the immigration interval varies. We made highly cooperative societies from initial networks of $N_i = 20000$, $30000$, and $40000$. We implemented $10$ different societies per each $N_i$. Each society has a different number of natives ($N$). In these societies, we set $R_c=0.8$ and the number of immigrants that immigrate at one time is $1\%$ of $N$. The results are shown in Fig.~\ref{fig2}(a). Average values of $N$ are $4848$, $7366$, and $9685$ for $N_i=20000$, $30000$, and $40000$, respectively. Since $\chi$ values are almost independent on the population of the society, the maximum number of immigrants for successful immigration is proportional to the population of the society. For identifying other properties of immigration, we make highly cooperative society with $N_i=30000$ and $N = 7582$. Also, we fix the number of immigrants at one time to $50$ from now on. In this condition, we observe $\chi$ with varying $R_c$ and $R_i$. The results are shown in Fig.~\ref{fig2}(b). When total number of immigrants $R_i$ is small and $R_c$ is close to $1$, the immigration is successful ($\chi > 0$). Note that $\chi > 0$ if $R_c$ is very close to $1$ even when $R_i$ is large. To summarize the results in Fig.~\ref{fig2}, the population of the society before immigration occurs and the attitude (strategy) of immigrants are main factors for successful immigration.

Figure~\ref{fig3} shows the effect of the immigration interval when $R_c$ and $R_i$ are fixed. For $R_c=0.7$, $\chi$ does not depend on the immigration interval. However, for larger $R_c$, $\chi$ increases with larger immigration interval. When immigrants enter the society, the society undergoes a disturbed time, after which the society returns to a highly cooperative society. Thus, $\chi$ is saturated when the immigration interval is larger than the disturbed time. To increase the total number of immigrants while maintaining $\chi > 0$, the society needs enough immigration interval.

Now, we consider an extreme case. We suppose that all immigrants are cooperative. Since immigration does not introduce a defective strategy in the society, $\chi$ is always $1$ and the population of the society can grow infinitely. The question is whether the society changes its structure qualitatively with respect to the stability by the cooperative immigrants. In this paper, we define that the society is stable if the society maintains positive $\chi$ when $20$ defective invaders enter the society.

\begin{figure}
\centering
\includegraphics[angle=270,width=1\columnwidth]{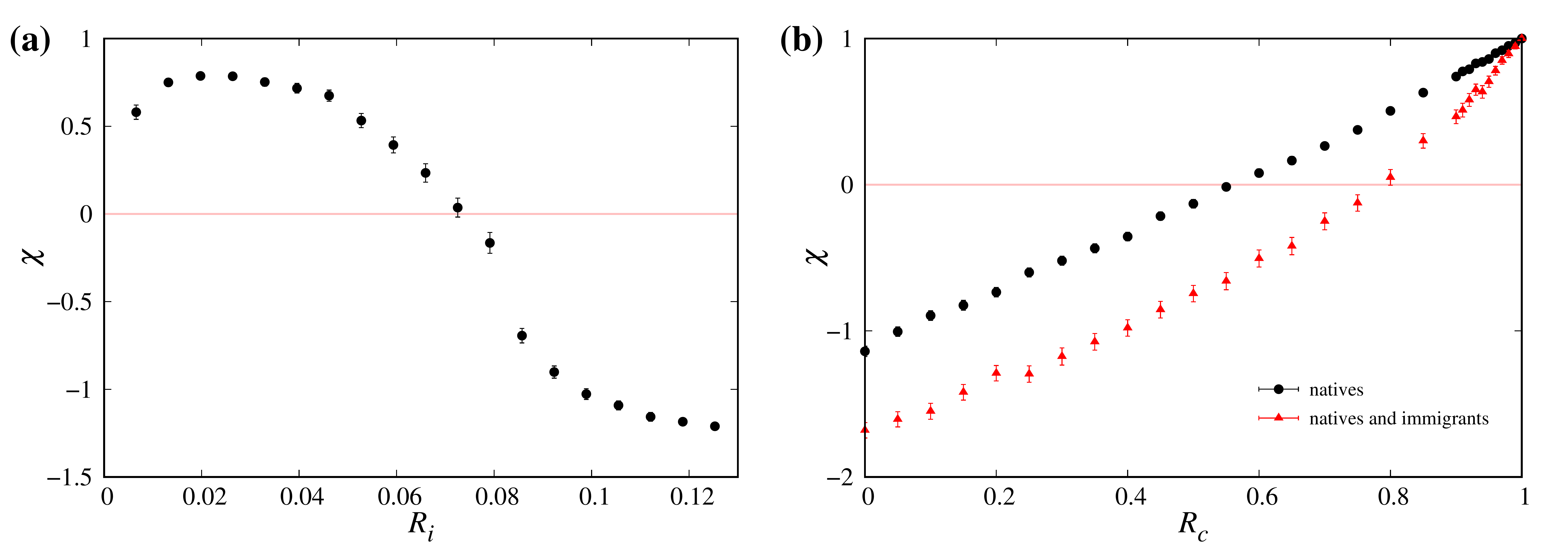}
\caption{We use the same society as in Fig.~2(b). (a)~$\chi$ by the invasion for each society, which has different $R_i$ and only cooperative immigrants. $1000$ simulations were averaged in each case. When $R_i$ is larger than $0.075$, $\chi$ is negative. (b)~$\chi$ as a function of $R_c$ for two highly cooperative societies with and without immigrants. The total number of immigrants is $200$ ($R_i \approx 0.02$) and the population is $7582$ in both societies. $1000$ simulations were averaged in each case.}\label{fig4}
\end{figure}

To verify the stability of the society, we immigrate only cooperative immigrants into the society until satisfying $R_i$ we want. After that, $10$ defective invaders enter the society (invasion). Next, we simulate $1000$ time steps, and then $10$ defective invaders enter the society once again. (Invasion interval is $1000$.) After the second invasion, we simulate $20000$ time steps more for adaptation of invaders. The results of the invasion are shown in Fig.~\ref{fig4}(a). When $R_i$ is small, $\chi > 0$. However, although the number of defective invaders is very small compared to the population of the society, $\chi$ is negative by invasion when $R_i$ is large. It is because immigration makes the structure of the society unstable. To verify this, we made two kinds of highly cooperative societies of $N=7582$. The first one has only natives and the second one is composed of $7388$ natives and $194$ immigrants. (The immigration in the second society occurred in four times: three times of $50$ immigrants and once in $44$ immigrants.) Figure~4(b) compares values of $\chi$ by $200$ additional immigrants in the two societies as a function of $R_c$. In the same $R_c$, $\chi$ of the society that is made up of natives and immigrants is less than $\chi$ of the society that is made up of only natives except for $R_c = 1$. The difference of $\chi$ increases as the ratio of the immigrants in the society increases. The society that is made up of natives and immigrants is vulnerable to the immigration of defectors. Thus, even if all immigrants have a cooperative strategy, the society may have to restrict the number of immigrants for the stability of the society.

Until now, we assumed that immigrants connect randomly to members of the society \cite{Tarik}. This approach is reasonable when immigrants have no information about the society. However, what if they have some information about members of the society? In the real world, immigrants may gather some information about the society and prepare for future treats. To study this effect, we adopt preferential attachment \cite{PhysRevE.73.056124,barabasi1999emergence}. Preferential attachment uses the degree of each member to be regarded as the member's reputation in the society. A person who has a high reputation is considered successful in the society, and immigrants want to have a relationship with the person. We simulate in the same society as in Fig.~\ref{fig2}(b), but each immigrant has one preferential attachment link. The other links are connected randomly as before.

\begin{figure}
\centering
\includegraphics[angle=270,width=0.46\columnwidth]{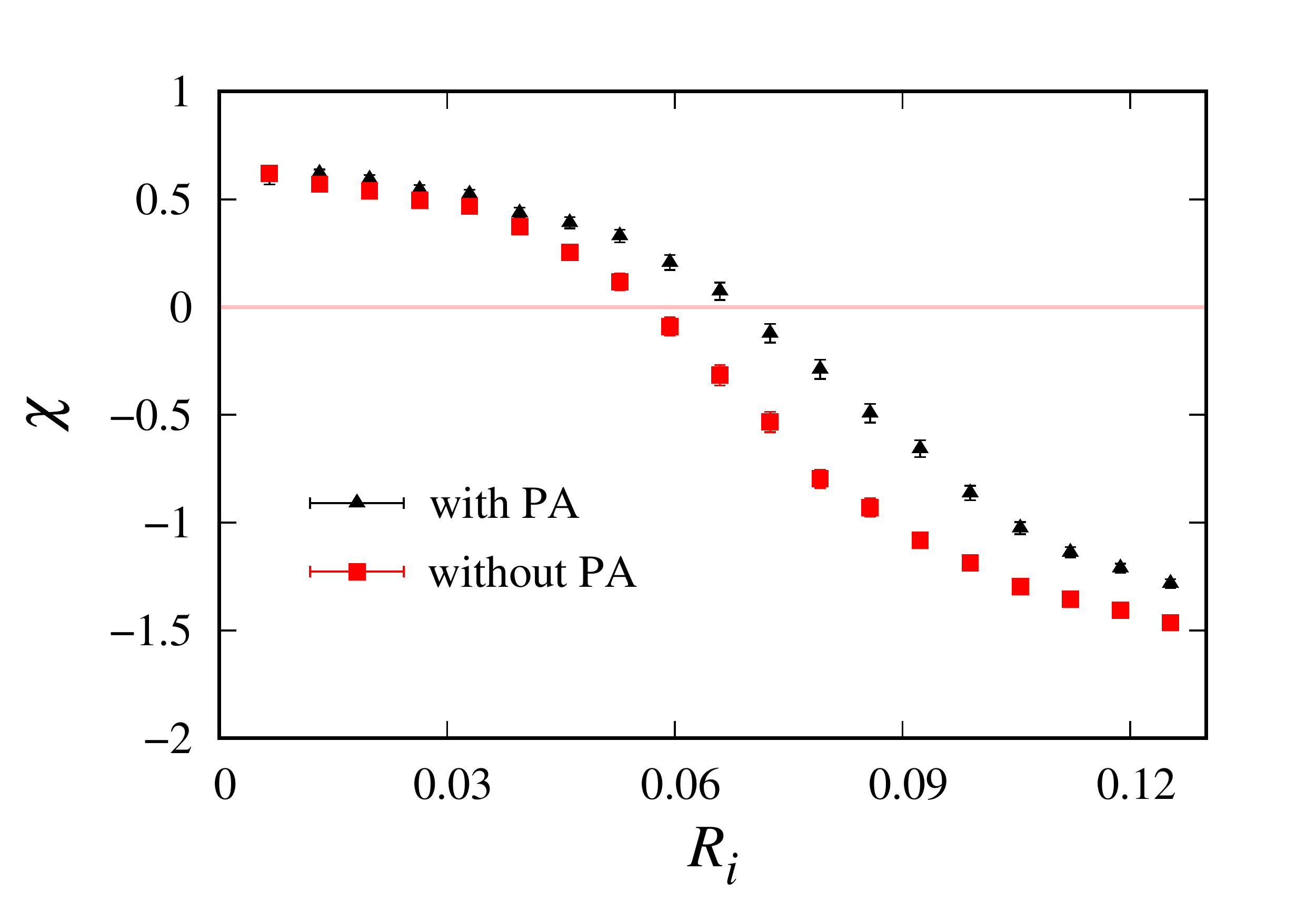}
\caption{Plot of $\chi$ as a function of $R_i$ with and without the preferential attachment (PA). The society before immigration is the same as in Fig.~2(b) and $R_c = 0.8$. Preferential attachment increases the number of acceptable immigrants while maintaining $\chi > 0 $. $1000$ simulations were averaged in each case.}\label{fig5}
\end{figure}

As shown in Fig.~\ref{fig5}, when immigrants have one preferential attachment, the number of acceptable immigrants while maintaining $\chi > 0 $ increases substantially. It is because immigrants can accept a suitable strategy in the society through a neighbor connected by the preferential attachment.

\section{Conclusions}

We have observed the effects of minimum requirements and immigration in the prisoner's dilemma game. Minimum requirements condition induces a highly cooperative society. When immigration occurs in this highly cooperative society, the number of immigrants that the society can accept while maintaining successful immigration depends on the population of the society, the ratio of cooperator among immigrants, and the immigration interval. In addition, changing the connection rule of immigrant's link from a random connection to a preferential attachment increases it. Additionally, even if all immigrants are cooperative, the excessive acceptance of immigrants makes the society unstable.

\section*{Acknowledgments}
This work was supported by GIST Research Institute (GRI) grant funded by the GIST in 2018.


\bibliography{project1}

\end{document}